\documentclass[12pt]{JHEP}
\usepackage{graphics}
\usepackage{cite}
\newcommand{\beq}{\begin{equation}}
\newcommand{\eeq}{\end{equation}}
\newcommand{\bea}{\begin{eqnarray}}
\newcommand{\eea}{\end{eqnarray}}

\renewcommand{\b}{\beta}

\newcommand{\s}{\sigma}

\newcommand{\rf}[1]{(\ref{#1})}
\newcommand{\ra}{\rightarrow}

\title{Broadening of the \boldmath QCD${}_3$ flux tube from the AdS/CFT 
correspondence} 

\author{Jeff Greensite \\ Physics and Astronomy Dept.,
San Francisco State University \\
San Francisco, CA 94132 USA \\
E-mail: \email{greensit@quark.sfsu.edu}}

\author{Poul Olesen \\ The Niels Bohr Institute,
Blegdamsvej 17, DK-2100 Copenhagen \O, Denmark \\
E-mail: \email{polesen@nbi.dk} }

\abstract{We use the finite temperature AdS/CFT approach to
demonstrate logarithmic broadening of the confining $QCD_3$ flux tube
as a function of quark separation.  This behavior indicates that, unlike
lattice QCD, there is no roughening transition in the AdS/CFT
formulation, which raises the interesting possibility of extrapolating
strong coupling results to weak couplings by the use of resummation
techniques. In the zero-temperature non-confining limit, we find that
this logarithmic broadening of the field strength distribution is
absent.  Our results are obtained numerically at strong couplings, in
the supergravity approximation.}
  
\keywords{Confinement, AdS-CFT Correspondence, Brane Dynamics in Gauge
Theories}

\preprint

\begin{document}

\section{Introduction}

   Until recently, the analytical investigation of QCD at large distance
scales was only possible in the framework of strong coupling lattice
gauge theory.  This formulation has many beautiful features, and the
existence of confinement, chiral symmetry breaking, and a mass gap 
can be demonstrated in an elegant way.  It was once hoped that these
strong coupling results could be extrapolated to weaker couplings
via Pade approximants, or by some other clever resummation method.
These hopes were abandoned with the recognition that there exists
a roughening transition in lattice gauge theory, which separates the
strong and weak coupling phases.  In the roughened phase, the QCD
flux tube is believed (on the basis of some simple string 
arguments \cite{l})
to broaden as the quark-antiquark pair are separated; the cross-sectional
area of the flux tube should increase logarithmically with separation.
In the strong-coupling phase, however, this broadening simply does not
occur; the chromo-electric flux tube does not really behave like a
quantized string.  This unrealistic feature of strong-coupling lattice
theory is a fundamental limitation, which prevents the use of strong
coupling results to draw conclusions about physics in the continuum.

   The AdS/CFT correspondence put forward by Maldacena \cite{1}
provides a new way of doing strong coupling calculations in large-N
gauge theories with an ultraviolet cutoff. The effective cutoff is
provided by a compactified Euclidean time variable, equivalent to a
finite temperature, which supplies the required supersymmetry breaking
\cite{2}.  As in lattice gauge theory, the existence of confinement
and a mass gap can be elegantly demonstrated at strong couplings.  On
the other hand, as discussed by Gross and Ooguri \cite{3}, the
non-supersymmetric theory constructed in this way is really only
equivalent to large $N$ QCD in the limit where the temperature $T$
goes to infinity and the coupling $\lambda=g^2_{YM}N$ approaches zero,
with (in D=4 dimensions) \cite{3}
\beq
T\rightarrow\infty~~{\rm and}~~\lambda\rightarrow\frac{B}{\ln (T/\Lambda_
{QCD})},
\eeq
where $\Lambda_{QCD}$ is the QCD scale.
The supergravity approximation clearly breaks down
in these limits, where no results are presently available.  The AdS/CFT
approach is so far only tractable in the supergravity approximation,
where the temperature $T$ is an ultraviolet cutoff and 
$\lambda \gg 1$ 
is the bare coupling at the scale $T$. Only in this strong coupling limit 
has the string tension has been computed, via a saddle-point approximation 
\cite{tension}, and the low-lying glueball spectrum obtained. 
Some preliminary results in the one loop approximation
are also available \cite{loop1,loop2,loop3}.

   Since we are limited to strong couplings, it is interesting to ask
how far the results extracted from the AdS/CFT correspondence agree
with our expectations of large-N QCD in the continuum.  Further, we
would like to know if there is any fundamental obstruction, as there
is in lattice gauge theory, to extrapolating strong coupling AdS/CFT
results to weaker couplings.  In this article we will study one aspect
of these issues, namely, the question of whether strongly-coupled
QCD${}_3$ is in the roughened phase in the AdS/CFT correspondence.
The theory is in the rough phase, as explained many years ago by
L\"{u}scher, M\"{u}nster, and Weisz \cite{l}, if the QCD flux tube
broadens logarithmically with quark separation.  If this logarithmic
broadening, due to quantum vibrations of the flux tube, is \emph{not}
found at strong couplings, then there is likely to be a roughening
transition between the strong and weak coupling regimes.  As a
consequence, information about the strong coupling phase could not be
used to gain any insight in the physical weak coupling scaling
limit.

   Although the interquark potential is derived from a string theory
in the AdS/CFT correspondence, the question of flux tube broadening by
vibrations is non-trivial, since the potential is represented by a
string theory in both the supersymmetric and non-supersymmetric gauge
theories, while only the latter case would be expected to show
logarithmic broadening.  Moreover, our investigation will be limited
to the saddle-point evaluation of loop-loop correlation functions, and
vibrations around the saddle-point will not be considered.  In view of
this, the success of the AdS/CFT approach is noteworthy: We find, in the finite
temperature case, that the width of the QCD flux tube already shows
roughening in the saddle-point approximation, with the (width)$^2$
proportional to the logarithm of the interquark distance, as expected
for {\it vibrating} strings \cite{l}, \cite{oa}. For the zero
temperature case, where one has a conformal theory, this logarithmic
broadening does not occur.
The latter result is achieved in the AdS/CFT approach by
a remarkable fine tuning of the relevant string worldsheet in the zero
temperature case, which of course is related to the absence of a horizon.   
Superficially the metric and the relevant
saddle-point configuration look very similar to the $T\neq 0$ case, but
they nevertheless manage to produce, by subtle effects, a very
different behavior.   Thus, in the finite temperature case, the type IIB
string in $D=10$ dimensions $(AdS_5\times S_5)$ is associated with a 
confining flux
tube of finite width in flat $D=3$ dimensional space, while in the
zero-temperature case the superstring in 10 dimensions does not give
rise to a flux tube of this kind.

  The presence of roughening in the finite temperature case is
interesting on several grounds.  First, it adds to the list of
successes of the strong-coupling AdS/CFT formulation, which include
the existence of a linear interquark potential, a L\"{u}scher $1/R$
correction to that potential, a mass gap, and a quasi-realistic
pattern of glueball mass-splittings.\footnote{The existence of the
L\"{u}scher $1/R$ term in the AdS/CFT correspondence now seems very
plausible, c.f.\ ref.\ \cite{loop2}, although, as discussed in this
reference, some ambiguities remain, and the precise coefficient of
this term is still uncertain.  The interpretation of the
coupling-independent $1/R$ term in the potential is also complicated
by the fact that such a term also arises, in the AdS/CFT
correspondence, in the zero-temperature non-confining limit
\cite{loop3,Tseytlin}.}  Second, it demonstrates that there is
unlikely to be a roughening transition between strong and weak
coupling phases, and hence no obstruction (at least from this source)
to extrapolating strong coupling expansions to weaker couplings by
resummation methods.  Finally, the AdS/CFT correspondence provides the
first actual derivation of roughening in a non-abelian gauge theory,
albeit at strong couplings. Previous arguments for roughening \cite{l}
were based on a string picture which was not derived from gauge
theory, while numerical lattice simulations of non-abelian gauge
theory have neither confirmed nor denied the existence of roughening.
The flux tube width has in fact been studied numerically, but the
relevant error bars are simply too large to draw definite
conclusions.\footnote{Roughening has been observed in numerical
simulations of one simple abelian model, namely, $Z_2$ lattice gauge
theory in $D=3$ dimensions \cite{Gliozzi}.  But there is no evidence
for or against this effect in any non-abelian model.}  All that can be
said is that existing Monte Carlo data is compatible with logarithmic
broadening of the flux tube, but it is also perfectly compatible with
a constant width for the QCD flux tube \cite{wup}.

\section{Equations of motion in the saddle-point approximation}

   The color-electric energy density ${\cal E}(x)$ of a flux tube
running between static quark-antiquark sources is given by
\beq
     {\cal E}(\mbox{\bf x}) \propto \langle q\overline{q}|
       \mbox{Tr}\vec{E}^2(\mbox{\bf x})|
    q\overline{q}\rangle - \langle 0|\mbox{Tr}\vec{E}^2|0\rangle
\eeq
where $|q\overline{q}\rangle$ is the (normalized) ground state of
the gluon field in the presence of the static $q\overline{q}$ color
sources, and $|0\rangle$ is the ground state in the absence of
static sources. If we are interested, in particular, in the energy
density of the spatial component of the color electric field parallel
to the axis of the flux tube, then this is given by the connected 
correlator of a large $R \times T$ Wilson loop (denoted $C_2$)
with $T \gg R$, and a 
Wilson loop around a much smaller loop $C_1$ 
\beq
 {\cal E}_{||}(\mbox{\bf x}) 
\propto { \langle \mbox{Tr}[U(C_1)] \mbox{Tr}[U(C_2)]\rangle
   -  \langle \mbox{Tr}[U(C_1)]\rangle \langle \mbox{Tr}[U(C_2)]\rangle
   \over \langle \mbox{Tr}[U(C_2)]\rangle}
\label{calE}
\eeq
where $C_1$ is parallel to the plane of the 
larger loop and centered at point $\mbox{\bf x},t=0$, and where
$\mbox{Tr}[U(C)]$ denotes the trace of the Wilson loop around contour $C$.
Usually we are interested in the width of the flux tube in the
plane equidistant between the static quark sources.  If the quark
and antiquark sources
are located at  spatial positions $(0,0,-R/2)$ and $(0,0,R/2)$, respectively, 
then the width $w_R$ of the $QCD_4$ flux tube can be defined, e.g., by the
quantity 
\beq
      w^2_R \equiv {\int dx_1 dx_2 ~ (x_1^2+x_2^2) {\cal E}_3(x_1,x_2,0)
             \over \int dx_1 dx_2 ~ {\cal E}_3(x_1,x_2,0) }
\eeq
The flux-tube width $w_R$ is not very sensitive to the shape of
loop $C_2$.  Instead of a rectangular $R\times T$ loop, one could
use instead a circular loop of radius $R$ in eq.\ \rf{calE}; this corresponds
physically to creation of a heavy quark-antiquark pair which move 
gradually apart
to a maximum separation $R$, and then come gradually together again 
and annhilate.
If the small loop $C_1$ is concentric with loop $C_2$, at a transverse
separation $H$ as shown in Fig.\ \ref{circles}, then the connected
loop correlator measures the color-electric 
energy density of the flux tube at maximum quark separation $R$, at
a transverse distance $H$ from the midpoint of the axis of the flux tube.

  The quantity $w^2_R$ can be 
computed from a lattice strong coupling expansion.  At strong couplings,
$w^2_R$ goes to a finite constant in the $R\ra \infty$ limit;
i.e.\ the width of the flux tube is independent of the
quark-antiquark separation at large $R$. 
However, it was shown many years ago 
in ref.\ \cite{l} that $w^2_\infty$ diverges at $\b \approx 1.9$, for
SU(2) lattice gauge theory in $D=4$ dimensions.  
This is the roughening transition point, 
analogous to roughening transitions found in other statistical
systems.  An example of the phenomenon is the divergent 
surface fluctuations of a magnetic domain wall, at the roughening
transition point, found in the 3D Ising model.  It is the existence
of this phase transition in lattice QCD 
that prevents the extrapolation of strong-coupling
lattice calculations into the weak-coupling regime.

  Beyond the roughening transition, the width of the QCD flux tube
must grow without limit as quark separation $R$ increases.
L\"{u}scher, M\"{u}nster, and 
Weisz suggested \cite{l} that the two-Wilson loop
correlation function in eq.\ \rf{calE} might be represented by
the loop-loop correlation function $G(H;C_1,C_2)$ of a Nambu-Goto
string, which could be evaluated in the saddlepoint approximation.
For this calculation it is convenient to choose $C_1,~C_2$ to be 
concentric circles in parallel planes, of radii
$R_1$ and $R_2 > R_1$ respectively, separated by tranverse distance $H$  
(as already noted, the loops used in eq.\ \rf{calE} need not be 
rectangular). 

  In ref.\ \cite{l}, the calculation of $G(H;C_1,C_2)$ 
was done in flat space, and it was found that the cross-sectional
area of the flux tube, evaluated in saddlepoint approximation, grows
logarithmically with quark separation.  Of course, this result is
not decisive for QCD, because the relationship of the QCD 
flux tube to the Nambu-Goto
string is not entirely clear.  However, from the AdS/CFT 
correspondence, we now understand
the correct procedure (at strong couplings, large $N$, and
$D=3$ dimensions) to be as
follows: In an $AdS_5\times S_5$ background described by the
black-hole metric \cite{1,2,3,tension} appropriate to $QCD_3$, we
have
\beq
ds^2=\alpha'\left[{U^2\over R^2}\left(f(U)dt^2+dr^2+dz^2+r^2d\theta^2\right)+
\frac{R^2}{U^2}\frac{dU^2}{f(U)}+R^2d\Omega_5^2\right],
\eeq
with
\beq
f(U)=1-\frac{U_T^4}{U^4}, ~~R^2=\sqrt{4\pi g N}.
\eeq
The two loops shown in Fig.\ \ref{circles} are located at $U=\infty$, and at
a fixed point on $S_5$.  Each circular loop is centered at $r=0$ in 
a plane of constant $z=z_1,~z_2$ respectively, with $H=|z_1-z_2|$.
The $QCD_3$ correlation function for two Wilson loops, appearing
in eq. \rf{calE} is then determined by
the loop-loop correlation function $G(H;C_1,C_2)$
(i.e. the off-shell string propagator) 
in the $AdS_5\times S_5$ background, for the given loops 
at $U=\infty$.

\FIGURE[h]{
\centerline{\scalebox{0.5}{\includegraphics{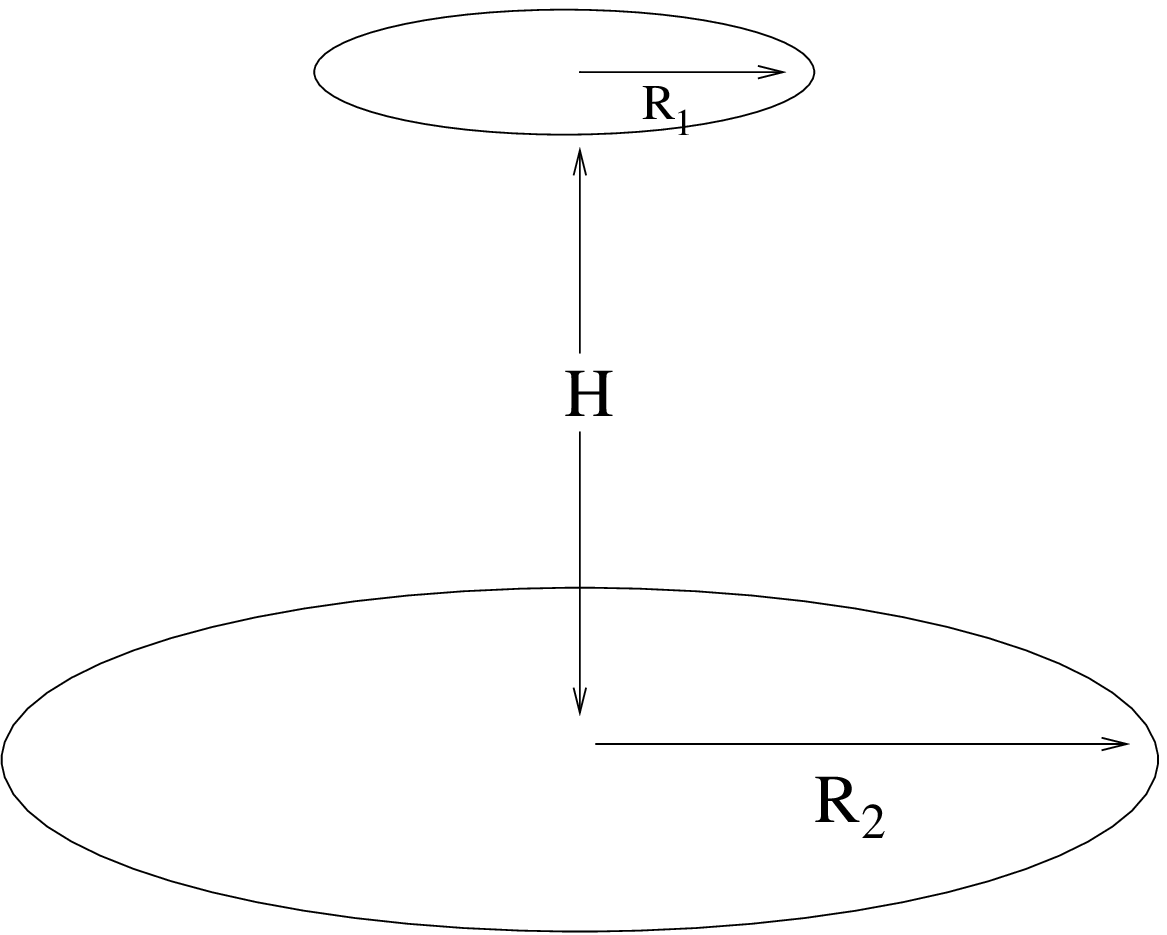}}}
\caption{Loops $C_1,~C_2$ in the Wilson loop correlator
$G(H;C_1,C_2)$.}
\label{circles}
}    

   Off-shell string propagators are complicated objects, even for
strings on a flat background \cite{Pol,Chaud}, but it will be sufficient 
for our purposes to work in the saddle-point approximation.  
This means that
\beq
G(H;C_1,C_2)\approx \exp[-S(H;C_1,C_2)],
\eeq
where $S$ is the worldsheet action of the extremal surface in the AdS
background, bounded by the two loops at $U=\infty$.\footnote{In the
zero temperature case, and for $R_1=R_2$, this propagator was computed
by Zarembo \cite{Zar} in the saddle-point approximation.}  The
saddle-point approximation has one important limitation: If $H$ is too
large, compared to the smaller of the two radii $R_1$, then the
minimal surface somewhere degenerates to a line.  At that stage the
semiclassical approximation has broken down, and a full quantum
treatment of the string worldsheet is required, as discussed in ref.\
\cite{3}.  However, we will see that the small $H$ behavior is
sufficient to observe logarithmic broadening.

  It is convenient to write the loop radii, transverse separation, and
$U_T$ as multiples of $R$
\beq
       R_1 = R L_1 ~,~~ R_2 = R L_2 ~,~~ H = R h ~,~~ U_T = R b
\eeq
and also to rescale coordinates by the corresponding substitutions
\beq
       U \ra R U ~,~~ r \ra R r ~,~~ z \ra R z
\eeq 
We choose to parametrize surfaces by the rescaled metric 
coordinates $r,\theta$,
and the symmetries of the problem allow us to consider surfaces with
AdS coordinates $U(r),z(r)$ independent of $\theta$. 
The worldsheet action in the $AdS_5\times S_5$ background is easily derived,
\beq
S=R^2\int_{r_0}^{L_2} dr~rU^2\sqrt{F},
\label{action}
\eeq
with
\beq
F=1+\left(\frac{dz}{dr}\right)^2+\frac{1}{U^4-b^4}\left(\frac{dU}{dr}\right)^2.
\label{F}
\eeq
The lower limit of integration in \rf{action} is determined by the
boundary conditions, as we will see in the next section.
The action (\ref{action}) does not contain the variable $z$ explicitly, and
hence we obtain the integral
\beq
\frac{dz}{dr}=\frac{q\sqrt{F}}{rU^2}.
\label{z}
\eeq
Here $q$ is an integration constant. By use of (\ref{F}) we then obtain
\beq
\left(\frac{dz}{dr}\right)^2=\frac{q^2~[1+(U^4-b^4)\phi^2]}{r^2U^4-q^2},
\label{zeq}
\eeq
where we introduced the notation
\beq
\phi=\frac{1}{U^4-b^4}~\frac{dU}{dr}.
\label{phi}
\eeq
The quantity $F$ in eq. (\ref{F}) can then be reexpressed as
\beq
F=\frac{r^2U^4(1+(U^4-b^4)\phi^2)}{r^2U^4-q^2}.
\label{FF}
\eeq
We also need the second derivative of $z$, which is most easily obtained
from the hamiltonian $H$ constucted from the action (\ref{action}), with
$dH/dr=-\partial {\cal L}/\partial r$. In this way we get
\beq
\frac{d}{dr}\left(\frac{rU^2}{\sqrt{F}}\right)=U^2\sqrt{F}.
\label{motion}
\eeq
Inserting eq. (\ref{z}) this leads to
\beq
\frac{d^2z}{dr^2}=-\frac{rU^4}{q^2}\left(\frac{dz}{dr}\right)^3.
\label{doublez}
\eeq

We can then derive an equation of motion for the field $\phi$. Starting from
eq. (\ref{motion}) we obtain after a tedious calculation
\beq
r\frac{d\phi}{dr}=\left(\frac{2r}{U}-\phi\right)\frac{r^2U^4+q^2
(U^4-b^4)\phi^2}{r^2U^4-q^2}-\phi^2\left(\frac{2rb^4}{U}+(U^4-b^4)\phi\right).
\label{good}
\eeq 
Together with eqs.\ (\ref{zeq}) and (\ref{phi}) this constitutes the 
first-order saddle-point equations for the extremal surface, to be solved
for $U(r)$ and $z(r)$ given appropriate 2-loop boundary conditions. 
Alternatively, one can use eq.\ \rf{doublez} and
\beq
  \left( {dU\over dr}\right)^2 = (U^4-b^4)
     \left[ \left({r^2 U^4\over q^2} - 1\right)\left({dz\over dr}\right)^2
          - 1 \right]
\eeq
which follows from eqs.\ \rf{zeq} and \rf{phi}.

\section{Numerical Solution for the Minimal Surface}

   It is possible to obtain two asymptotic solutions to the equations of
motion derived above. The boundary of the world sheet corresponds
to $U=\infty$. Let this occur at $r=L$, where $L$ is  
to be identified with $L_1$ or $L_2$. 
We find without difficulty that for $U$ very large
\beq
U\approx \frac{1}{\sqrt{2L(L- r)}}.
\label{asym1}
\eeq
It is, however, also possible for $U$ to approach a constant value, $U_0$ say,
for $r\rightarrow r_0$ in such a way that the derivative of $U$ goes to 
infinity in this point. This behavior corresponds to
\beq
U\approx U_0+{\rm const.}\sqrt{r-r_0}.
\label{asym2}
\eeq
Obviously this requires $r>r_0$.

   The significance of these asymptotic forms can be appreciated by
skipping ahead a little bit, and looking at a particular minimal
surface, found by solving eqs. \rf{zeq},\rf{phi},\rf{good} numerically, 
for parameters $L_1=2,~L_2=4,~h=2.9,~b=0.5$.  Solutions for $U(r)$
(for $U \le 6$)
and $z(r)$ are shown in Figs. \ref{U1} and \ref{z1}.  By rotating these
curves in $\theta$ (i.e. around the $U,z$ axes in the respective figures),
one obtains a projection of the minimal surface in $AdS_5$ 
onto the $r\theta U$ and $r\theta z$ hyperplanes.  
   
\FIGURE[h]{
\centerline{\scalebox{0.9}{\includegraphics{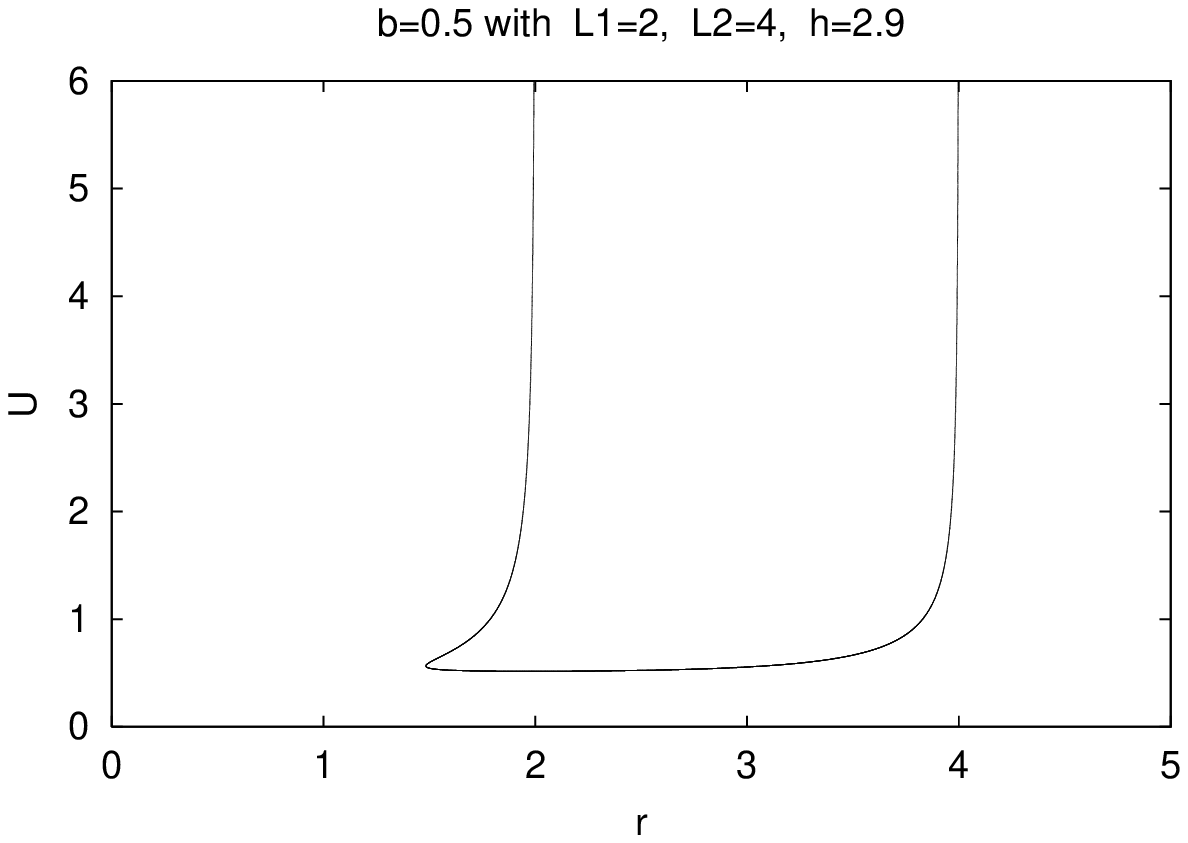}}}
\caption{Profile of $U(r)$. The double valuedness of this function should
be noticed.}
\label{U1}
}    

\FIGURE[h]{
\centerline{\scalebox{0.9}{\includegraphics{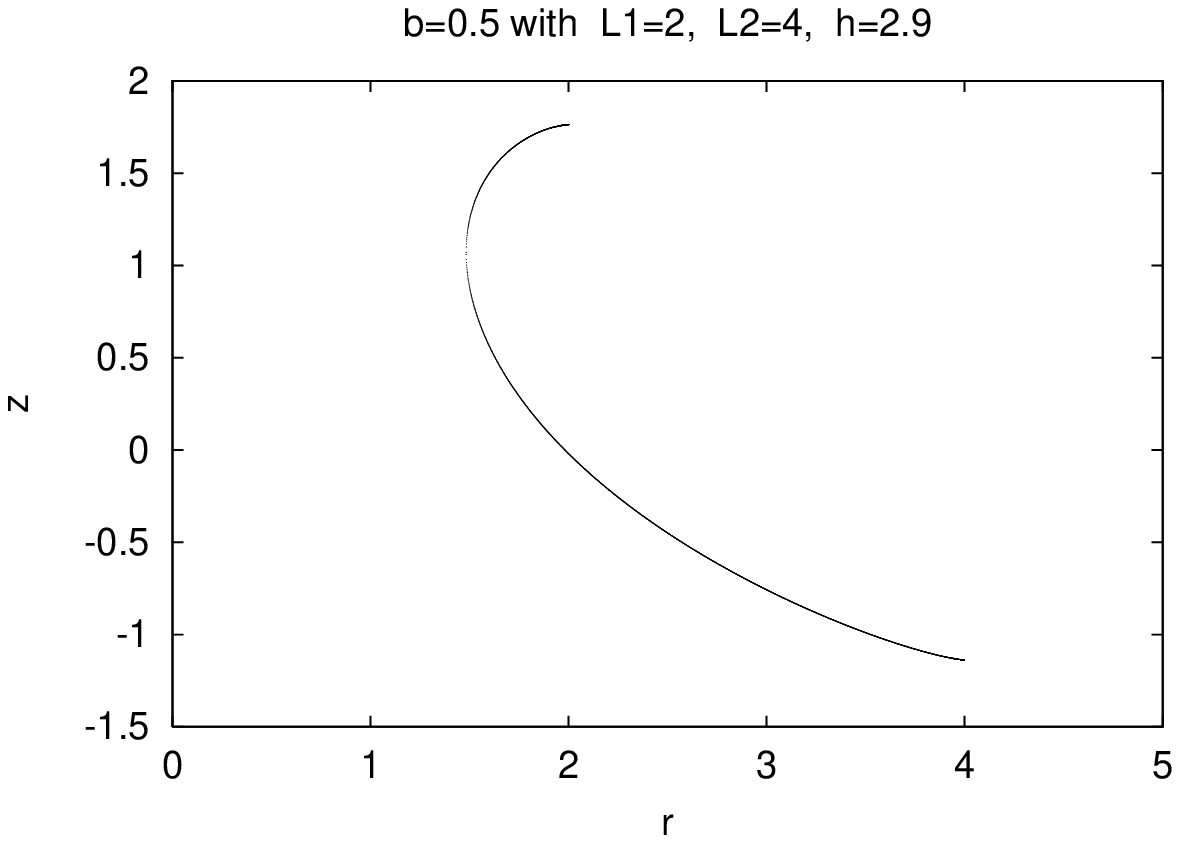}}}
\caption{A profile of the function $z(r)$, which is also double valued.}
\label{z1}
}    

   Note that at both $r \ra L_1$ and $r \ra L_2$, the slope $dU/dr$ is
positive, which can be immediately deduced from the asymptotic form
\rf{asym1}.  This means that $U(r)$ is double-valued in some finite
region $r_0<r<L_1$, and it turns out that $z(r)$ is also double-valued
in this range.  At $r=r_0$, the slope $dU/dr$ is infinite while
$U(r_0)$ is finite, and the form \rf{asym2} applies.

   Because of the double-valuedness of $U(r)$ and $z(r)$, as well as
for reasons of numerical stability, we have found it expedient to
solve the equations for the minimal surface numerically in the
following way: We begin by noting that a solution of the equations of
motion is completely determined by specifying the constant $q>0$, and
the coordinates $(r_m,U_m,z_m)$ where $U$ is a minimum ($\phi= 0$).
We can arbitrarily set $z_m=0$, since only the difference
$h=|z(L_1)-z(L_2)|$ is relevant.  Then we integrate the equations of
motion from $(r_m,U_m,z_m)$ in the positive $r$ direction, to the
limiting point $r = L'_2$ where $U\ra \infty$, and we denote
$z_2=z(L'_2)$.\footnote{Note that in practice, when we refer to the
condition that $U$ or its derivative is infinite, this is of course
taken to mean that $U$ or $dU/dr$ exceeds some large finite bound.}
In eq.\ \rf{zeq} for $dz/dr$, there is a sign ambiguity in taking the
square root (unless $q=h=0$).  This is resolved by (arbitrarily)
taking the slope to be negative (it does not pass through zero for
$r>r_m$.) Next, again starting from the initial point $(r_m,U_m,z_m)$,
we integrate in the negative $r$-direction until the point
$r_0,U_0,z_0$ is reached where $dU/dr \ra -\infty$ (also $dz/dr \ra
-\infty$ if $q \ne 0$).  Finally, resetting $dU/dr = +\infty$, we
integrate again in the positive direction from $r_0,U_0,z_0$ to the
limiting point $r=L'_1,z=z_1$ where $U=\infty$.  In this region the
sign of $dz/dr$ must be taken positive.  Thus the integration is done
separately in three regions $[r_m,L'_2],~ [r_0,r_m],~[r_0,L'_1]$, and
the double-valuedness problem is circumvented.  It is then only
necessary to choose parameters $r_m,U_m,q$ such that
$L'_1=L_1,~L'_2=L_2,~h=|z_1-z_2|$ for given $L_1,~L_2,~h,$ which is
achieved by a simple Newton-Raphson method.  We note in passing that
the example shown in Figs. \ref{U1}, \ref{z1} was chosen so that the
$r_0<L_1$ feature, and the double-valuedness of $U(r),z(r)$ in the
region $r\in [r_0,L_1]$, are very pronounced.  In this example we had
$h=2.9 > L_1=2$.  Generally, for the region of interest $h \ll L_1$,
it is found that $r_0$ is quite close to $L_1$, and the
double-valuedness of $U(r),z(r)$ is not so apparent.

   Having obtained a numerical solution for the extremal surface bounding
two loops, we can substitute the result back into the expression
for the worldsheet action, eq.\ \rf{action}, which is 
then integrated numerically.
Of course, the action is infinite if the loops are actually placed
at $U=\infty$, so we regulate this infinity by placing the boundary loops at
a finite $U_{max}=30$.  Then we can separate out an $h$-dependent
area contribution
\beq
       S[H;C_1,C_2] = R^2 \Bigl( A_0[L_1,L_2] + A[h;L_1,L_2] \Bigr)
\eeq
where $A[0;L_1,L_2]=0$. The term $R^2A_0[L_1,L_2]$ is the area of the string
worldsheet at $h=0$, which includes an irrelevant quark self-energy
term, divergent as $U_{max}\ra \infty$. We are mainly interested in
the second, $h$-dependent term.
    
   In flat space, the worldsheet action $S[H,C_1,C_2]$ for a minimal
surface between parallel concentric loops can be calculated analytically,
and the result, for \\ 
$h \ll L_1 \ll L_2$, is \cite{l}
\beq
       S_{flat}[H;C_1,C_2] = \s \pi R^2(L_2^2-L_1^2) + {H^2 \over d^2}
\label{Sflat}
\eeq
where
\beq
        d^2 = {1\over \pi \s} \ln{L_2\over L_1}
\label{d2}
\eeq
with $\s$ the string tension.
The logarithmic dependence of $d^2$ on $L_2$ is evidence, in the saddle-point
approximation, of logarithmic broadening of the flux tube.  
L\"{u}scher, M\"{u}nster, and Weisz also argued in ref.\ \cite{l} that
this logarithmic dependence on $L_2$ holds beyond the saddle-point
approximation, in the $L_1 \ra 0$ limit, where the factor of $L_1$
in eq.\ \rf{d2} is replaced by a short-wavelength cutoff $\lambda$.

   To check for this logarithmic broadening in the AdS/CFT correspondence,
we first fit the $h$-dependent part of the worldsheet action,
$A[h;L_1,L_2]$, to a parabola 
\beq
         A[h;L_1,L_2] \sim {h^2 \over d^2}
\eeq
so that, for small $h \ll L_1$ ($H \ll R_1$), we have
\beq
         G[H;C_1,C_2] \sim \exp[-H^2/d^2]
\eeq
where $d$ will depend on $L_1,~L_2$.
Of course this parabolic 
form will not hold for large $h$, where we would expect
not a gaussian but rather a simple exponential falloff for the loop-loop
correlation function
\beq
      G[H;C_1,C_2] \sim \exp[- m_G H]
\eeq
where $m_G$ is the lowest-lying glueball state, corresponding to 
a dilaton exchange in $AdS_5\times S_5$.  This asymptotic behavior
cannot be accurately obtained in the saddle-point approximation, which
breaks down for $h \gg L_1$, but even in the saddle-point approximation there is
a deviation from gaussian falloff at large enough $h$.  Therefore,
in fitting $A[h;L_1,L_2]$ to a parabola, we obtain $d$ from a fit
in the restricted range of $h\in [0,L_1/4]$.\footnote{All fits were obtained 
using the data fitting capabilities of the GNUPLOT software package.}

   At small $h$, our numerical solutions for
$A[h;L_1,L_2]$ are fit very accurately by a parabola, as shown
in Fig.\ \ref{a4_12}, where we plot our data for $A[h;4,12]$ (crosses)
versus a parabola (solid line) with $d=1.492$.  In general we can extract
$d$ from the numerical solution to at least three digit accuracy.  For larger 
$h \gg L_1/4$, the worldsheet action goes over to a linear increase
with $h$, as shown in Fig.\ \ref{area4_12} (the slope is proportional
to $L_1$).  At large enough $h$ the saddle-point equations have no solution, 
as already noted above.  

\FIGURE[h]{
\centerline{\scalebox{0.4}{\rotatebox{270}{\includegraphics{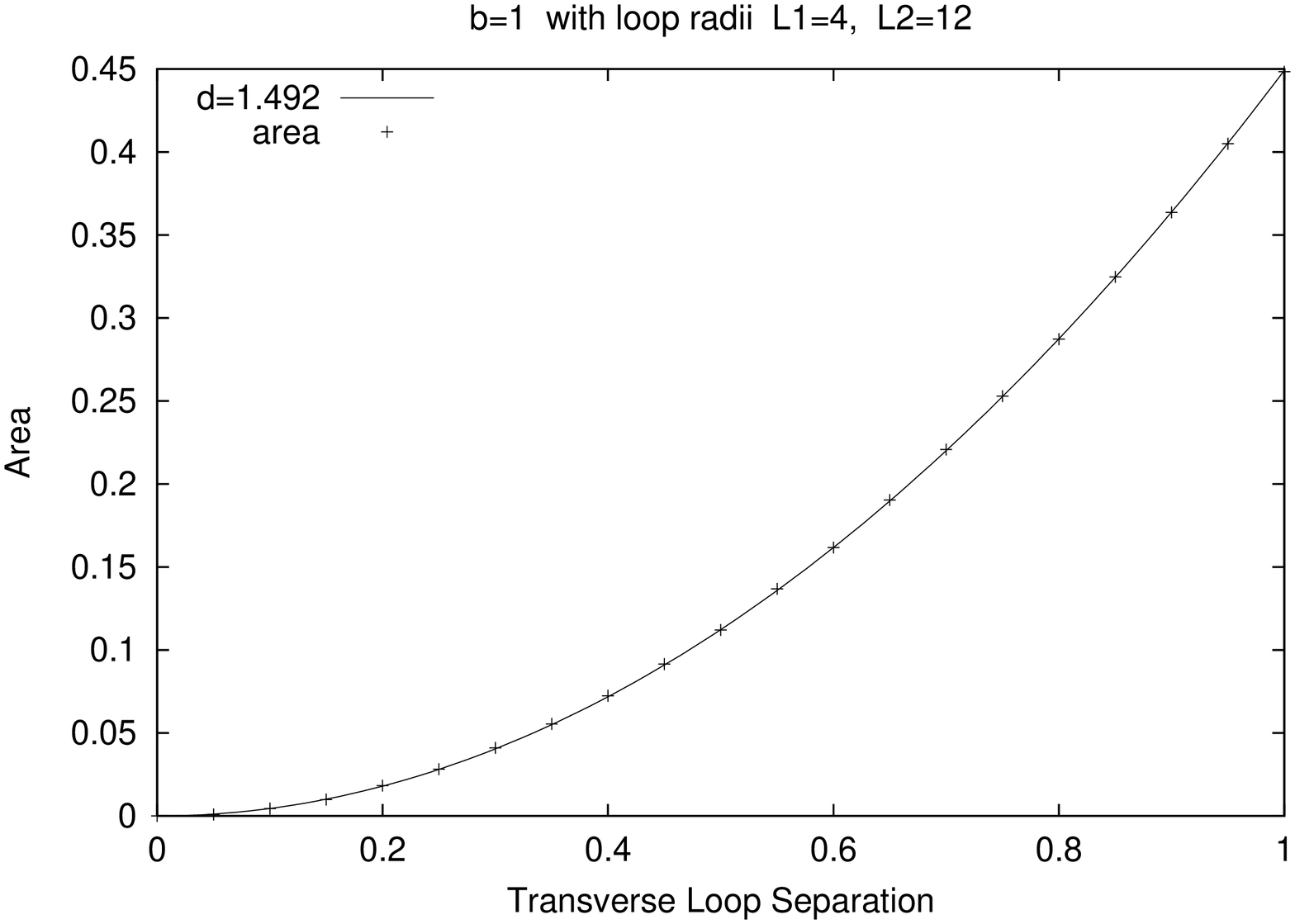}}}}
\caption{The $h$-dependent part of the worldsheet area $A[h;L_1,L_2]$
(denoted ``area'') at fixed radii, and the best parabolic fit.  Note
that $A[h;L_1,L_2]$ is the worldsheet ${\rm area}/R^2$ with a
subtraction, so as to equal zero at $h=0$.}
\label{a4_12}
}    

\FIGURE[h]{
\centerline{\scalebox{0.4}{\rotatebox{270}{\includegraphics{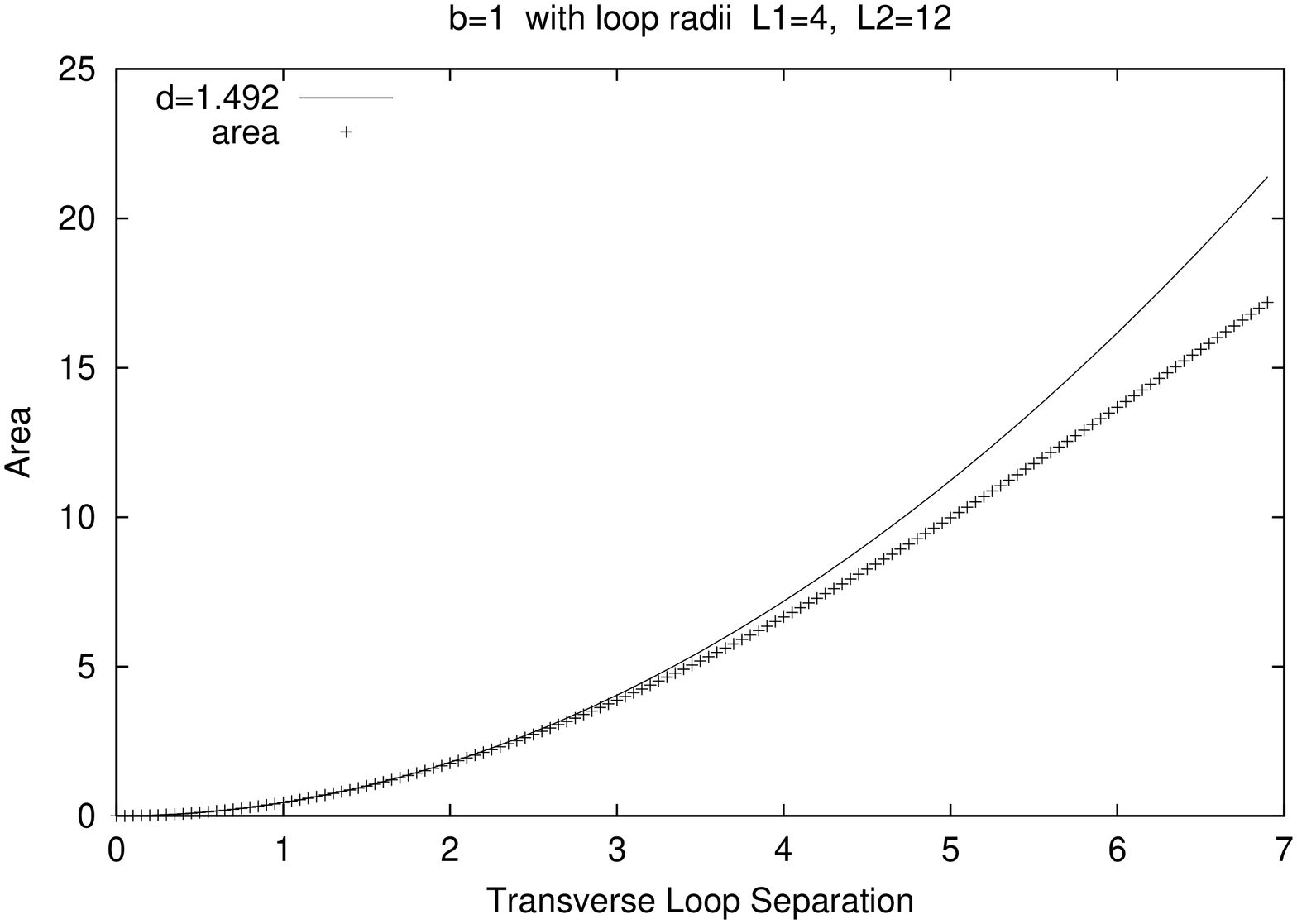}}}}
\caption{The $h$-dependent area term $A[h;L_1,L_2]$ (denoted ``area'')
at larger values of $h$. The parabolic fit obtained at $h\in [0,1]$ is
also displayed.  For $h> 3.5$ the behavior becomes linear in $h$.}
\label{area4_12}
}

  In AdS/CFT, the string tension at strong coupling is given by
\beq
       \s = {b^2 \over 2 \pi} = {U_T^2 \over 2\pi R^2}
\label{s_ads}
\eeq 
Therefore, if the L\"{u}scher, M\"{u}nster, and Weisz formula
\rf{d2} holds for the flux tube in the AdS/CFT correspondence, then
one expects
\beq
        d = {1\over b} \left[ 2 \ln{L_2\over L_1} \right]^{1/2}
\label{d}
\eeq

   In Fig.\ \ref{b1} we plot the tube radius $d$ vs.\ the ratio
$L_1/L_2$ for a variety of $L_1,L_2$ values, at fixed finite
temperature $b=1$.\footnote{Numerical instability limits our choice
of $L_1$ and $L_2$, since the allowable numerical error required
for results of fixed accuracy decreases
exponentially as $L_2$ increases.}  It appears that the numerical
solution for $d$ depends only on the ratio $L_1/L_2$, and fits eq.\
\rf{d} quite well. We are therefore justified in concluding that the
cross-section of the $QCD_3$ flux tube, as probed by the smaller loop
of radius $L_1$, broadens logarithmically with $L_2/L_1$ in
the strong-coupling AdS/CFT correspondence.  There is then no reason to
expect, in the AdS/CFT correspondence, a roughening transition
separating strong and weak-coupling phases. 

\FIGURE[h]{
\centerline{\scalebox{0.4}{\rotatebox{270}{\includegraphics{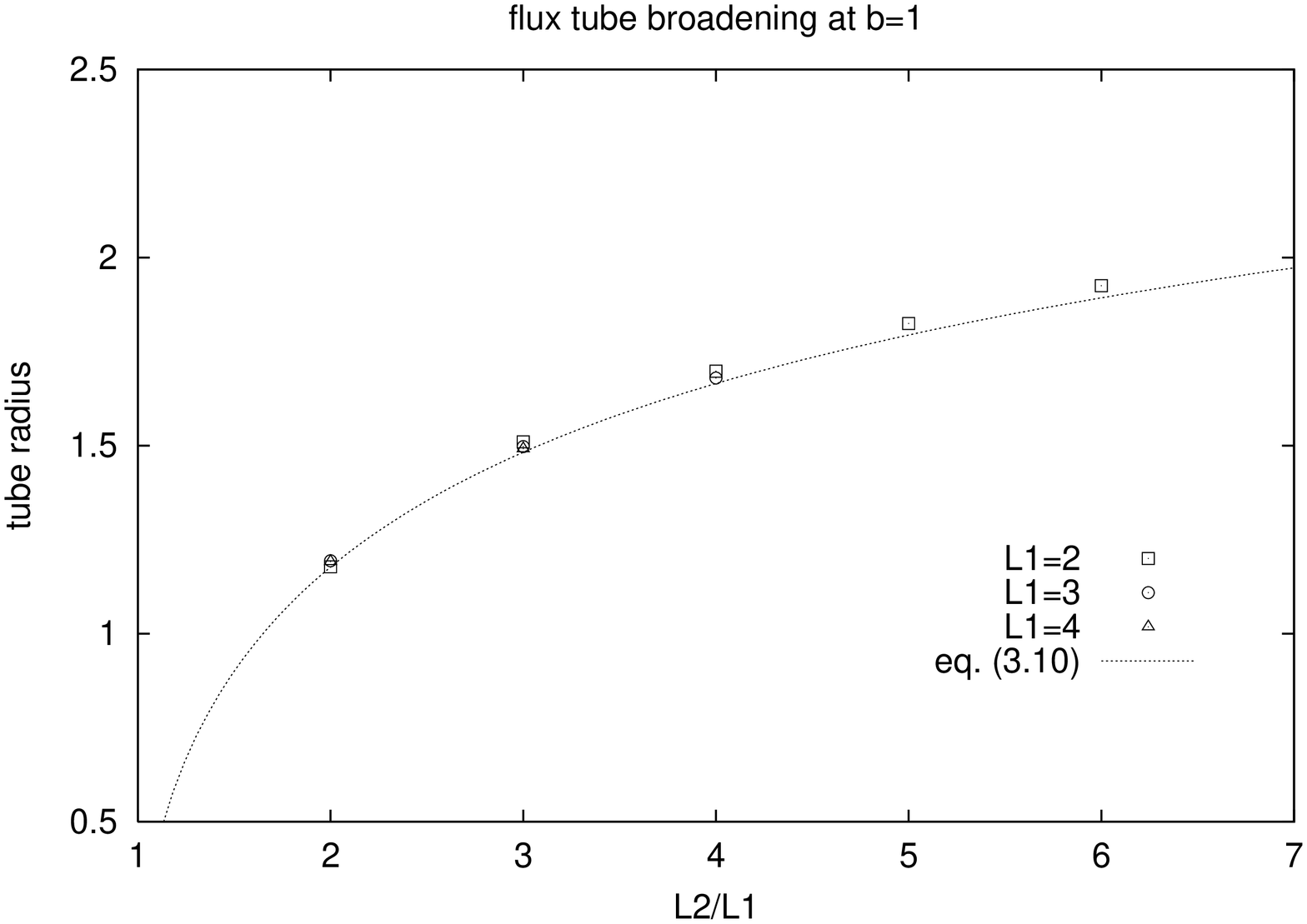}}}}
\caption{The tube radius $d$ as a function of $L_1/L_2$ with fixed 
temperature $b=1$.}
\label{b1}
}    

   Next we investigate the variation of $d$ with the temperature
parameter $b$.  According to eq.\ \rf{d}, $d$ varies inversely
with $b$, but this equation cannot possibly be right in the
$b\ra 0$ limit, where supersymmetry is restored and the confining
flux tube disappears.  For fixed $L_1=2,~L_2=4$, the variation
of $d$ with $b$ is shown in Fig.\ \ref{bv}.  The solid line is
eq.\ \rf{d}, which again fits the numerical solution very well
at these values of $L_1,~L_2$ for $b \ge 1$.  As $b \ra 0$,
however, $d$ goes to a finite constant.
   
\FIGURE[h]{
\centerline{\scalebox{0.4}{\rotatebox{270}{\includegraphics{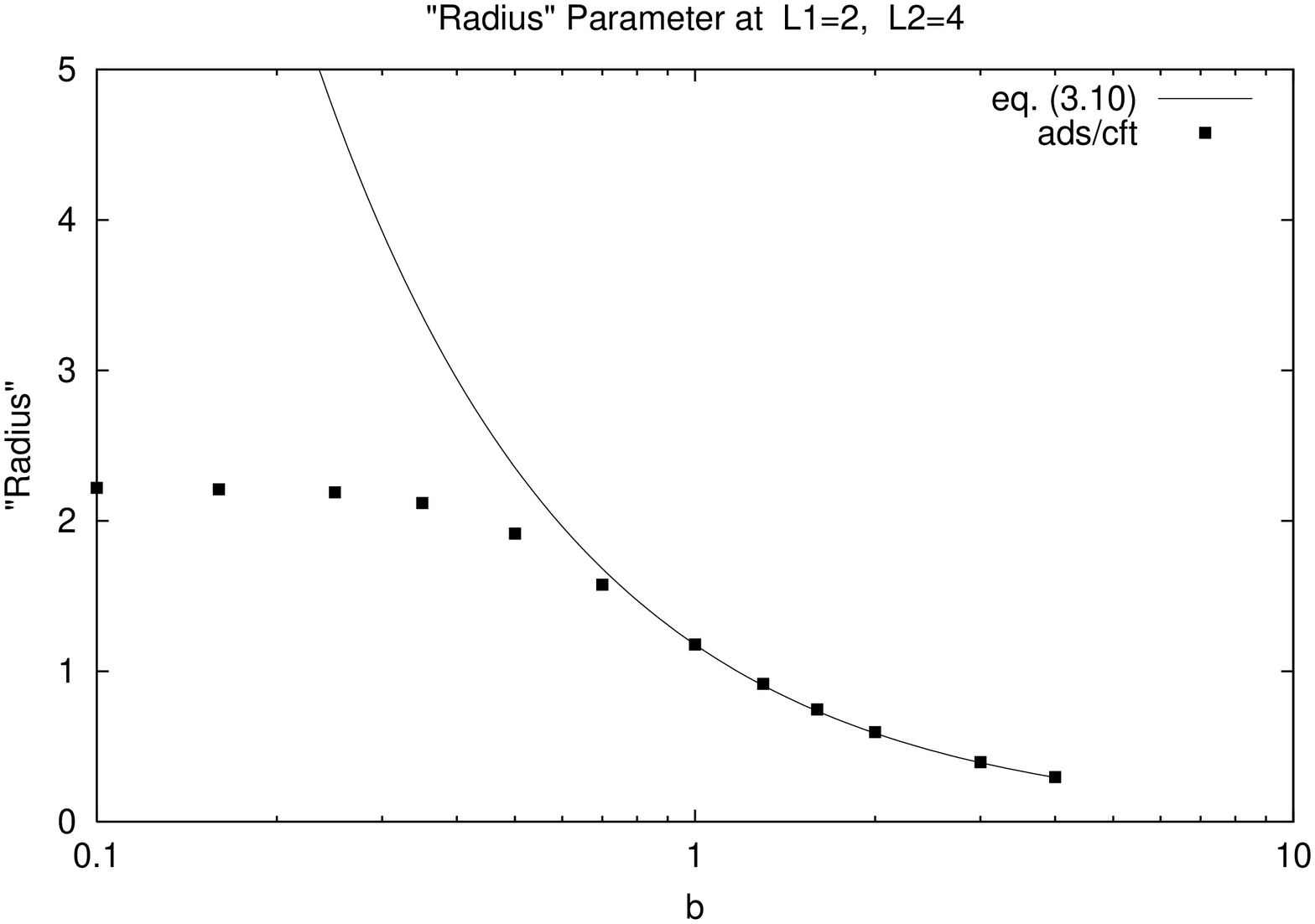}}}}
\caption{Variation of the ``radius'' $d$ with the temperature.}
\label{bv}
}

The result that $d$ ceases to depend on the logarithm of $L_2/L_1$ in 
the zero temperature case can be understood from a 
straightforward generalization of
the work by Zarembo \cite{Zar} (where he studied the case $L_1=L_2$)
to the case where $L_1$ and $L_2$ are different.  In the $L_1 \ne L_2$
case one finds, after expanding the action to lowest order in $h^2$ and
dropping the perimeter term \cite{pz}
\begin{equation}
S\approx R^2G(k_0)+\frac{G'(k_0)}{2(1-R_1^2/R_2^2)F'(k_0)}~\frac{H^2}{L_2^2},
\label{S}
\end{equation}
where $k_0$ is determined in terms of the ratio $R_2/R_1$ by the 
transcendental equation
\begin{equation}
F(k_0)=\frac{k_0}{2}\int_0^1 du\frac{\sqrt{\sqrt{1+4k_0^2(1-u)}-1}}
{\sqrt{u\left(2k_0^2+1-\sqrt{1+4k_0^2(1-u)}~\right)
\left(1+4k_0^2(1-u)\right)}} 
=\frac{1}{2}\ln \frac{R_2}{R_1}.
\label{trans}
\end{equation}
In eq. (\ref{S}) the quantity $F'(k_0)$ is the derivative of the function
$F$, and $G'(k_0)$ is the derivative of the function
\begin{equation}
G(k_0)=-\frac{2\alpha}{\sqrt{\alpha-1}}\int_0^{\pi/2}\frac{d\psi}{1+
\alpha \sin^2\psi+\sqrt{1+\alpha \sin^2\psi}},~~~~\alpha=\frac{1+2k_0^2+
\sqrt{1+4 k_0^2}}{2k_0^2}.
\end{equation}
To compare eq. (\ref{S}) with our numerical results we take $R_2/R_1=2$,
corresponding to $k_0\approx 2.53$, obtained by solving (\ref{trans}).
We then evaluate the derivatives entering in eq. (\ref{S}) numerically
and obtain
\begin{equation}
S(H)\approx 3.094 \left(\frac{H}{L_2}\right)^2,
\label{rrr}
\end{equation}
in good
agreement with the result shown in Fig.\ \ref{bv}. In this case the
width is given by
\begin{equation}
d\approx \frac{L_2}{\sqrt{3.1}}~~{\rm for}~~\frac{L_2}{L_1}=2.
\label{d_free}
\end{equation}
For different ratios $L_1/L_2$ the transcendental equation (\ref{trans})
should again be solved, the derivatives of $F$ and $G$ be computed and 
inserted in (\ref{S}). The coefficient of $H^2/L_2^2$ depends on 
$L_1/L_2$, so there is no universality in the coefficient. As an
example illustrating this, we can take $L_2/L_1=2.7$ 
(corresponding to $k_0\approx 0.70$), which 
gives $S(H)\approx 0.61 H^2/L_2^2$. The variation of the coefficient
of $H^2/L_2^2$ from $L_2/L_1=2.7$ to $L_2/L_1=2$ is $0.61/3.1\approx 0.20$. 
However, the corresponding logarithms would vary like $\ln 2/\ln 2.7\approx
0.70$, so the decrease of the coefficient with the ratio $L_2/L_1$ cannot be
explained by a logarithmic width as in the finite temperature case. It is 
also quite clear from Fig.\ \ref{bv} that the logarithmic fit fails for small 
temperatures, and $d$ does not have the interpretation of the radius of a 
flux tube. This is quite satisfactory, since there does not exist
any flux tube at zero temperature due to the vanishing string tension.

The parameter $d$ at $b=0$ can be extracted just as in the
finite temperature case.  In Fig.\ \ref{b0} we plot $d$ vs.\ $L_2$,
with $L_1/L_2 = \frac{1}{2}$ kept fixed.  We see that
the values obtained 
numerically for $d$ are in excellent agreement with the expected 
behavior (\ref{d_free}).\footnote{Correlation between large
and small loops in the zero temperature theory has also been discussed by
Berenstein et al.\ \cite{Ber}.  In that reference, the correlation is
calculated by running propagators (dilaton, graviton,...) between the
small loop and points on the classical worldsheet of the large loop.
The relation of their result to ours in not entirely direct, since
for $H\ll R_2$ and $R_1 \rightarrow 0$ one has to sum over propagators for
many excitations, and it is also not clear if, in this limit, the quantum 
fluctuations of the large loop worldsheet can be neglected.}

\FIGURE[h]{
\centerline{\scalebox{0.4}{\rotatebox{270}{\includegraphics{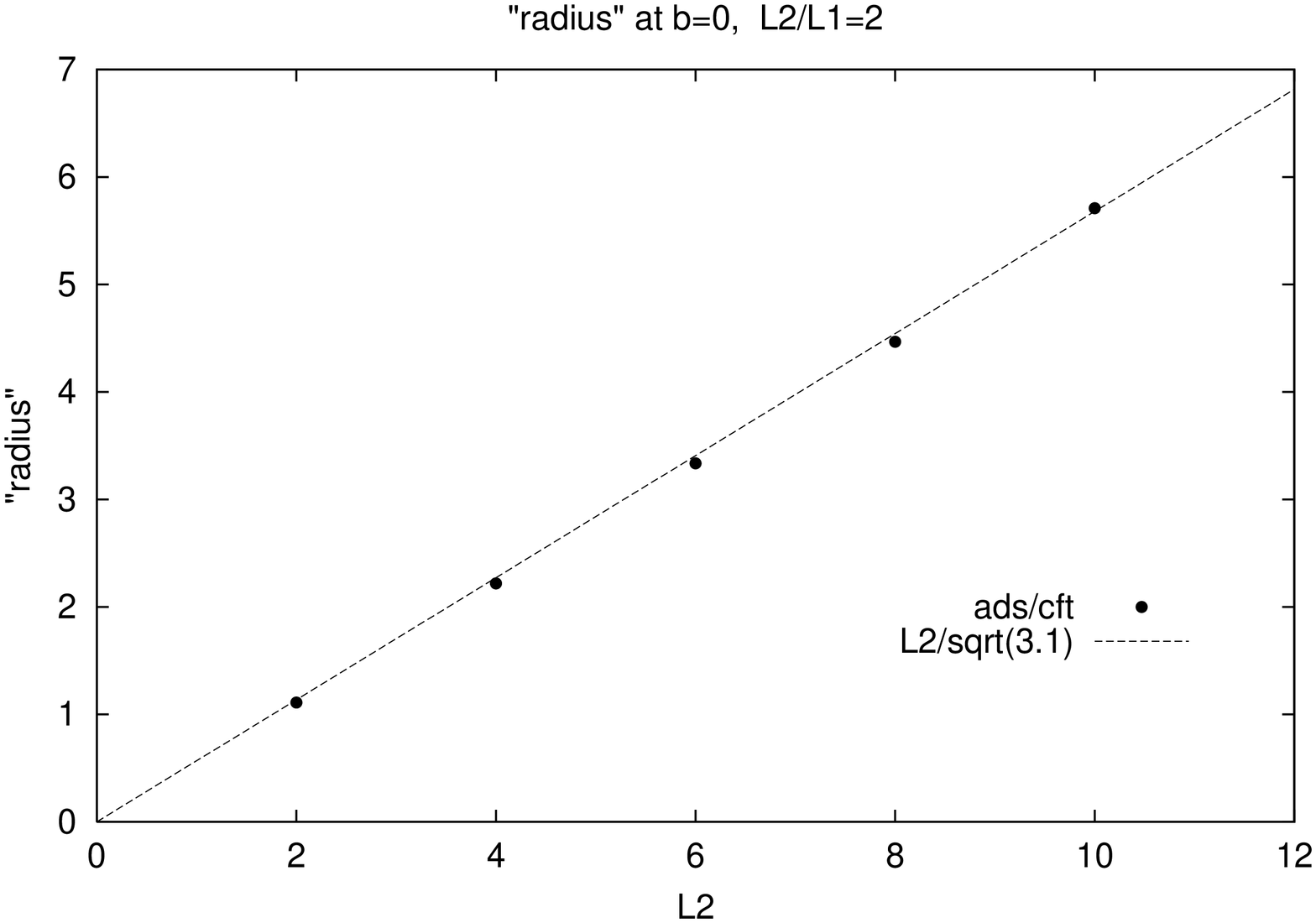}}}}
\caption{Zero temperature behavior of the radial parameter $d$.}
\label{b0}
}

\section{Analytic Approximation}

  From the numerical result we see that the behavior of $U$ is characterized 
by a ``plateau'' where $U$ is approximately a constant. If we now make
the somewhat primitive approximation that $U\approx U_{\rm min}$ and
the derivative $dU/dr\approx 0$ on the plateau, then from eqs. (\ref{phi}) and
(\ref{zeq}) we obtain for the height $h$
\bea
h&=&q\int_{{\rm whole~range~of}~r}dr\sqrt{\frac{1+(U^4-b^4)\phi^2}{r^2U^4-q^2}}
\approx q \int_{L_1}^{L_2} dr\frac{1}{\sqrt{r^2U_{\rm min}^4-q^2}}\nonumber \\
&=&\frac{q}{U_{\rm min}^2}\ln\frac{L_2+\sqrt{L_2^2-q^2/U_{\rm min}^4}}
{L_1+\sqrt{L_1^2-q^2/U_{\rm min}^4}}.
\label{analytic1}
\eea
In the second step we have ignored the contributions to the integral
away from the plateau.
These are essentially the irrelevant infinite self energy contributions.
Further, we
used that on the plateau one has that the quantity
\beq
P\equiv (U^4-b^4)\phi^2=\left(\frac{dU}{dr}\right)^2\frac{1}{U^4-b^4}
\eeq
is very small relative to one. This can be seen from the numerical
results,
which show that relative to one the quantity $P$ is at most of the order
of a few
per cent in a range going from slightly above $L_1$ to slightly below
$L_2$.
Near the minimum, $P$ behaves like
\beq
P\approx \frac{1}{2U_{\rm min}^3}\frac{d^2U}{dr^2}.
\eeq
On the plateau the second derivative of $U$ is very small, as can be
seen from the numerical results. 
For $L_1$ and $L_2$ sufficiently large relative to $q/U_{\rm min}^2$ we
obtain
\beq
\frac{U_{\rm min}^2}{q}\approx \frac{1}{h}\ln\frac{ L_2}{L_1}.
\label{pp}
\label{analytic2}
\eeq
Since $U_{\rm min}$ approaches $b$ with small exponential corrections for large
$L_1$ the quantity $q/U_{\rm min}^2$ is very close to $q/b^2$.

In the same approximation as in (\ref{analytic1}) we obtain for the action 
from eqs.(\ref{action}) and (\ref{FF})
\bea
S&=&R^2\int_{{\rm whole~range~of}~r}dr~r^2U^4\sqrt{\frac{1+(U^4-b^4)
\phi^2}{r^2U^4-q^2}}\approx R^2U_{\rm min}^4\int_{L_1}^{L_2}
dr \frac{r^2}{\sqrt{r^2U_{\rm min}^4-q^2}}\nonumber \\
&=&\frac{1}{2}~R^2U_{\rm min}^2\left[r\sqrt{r^2-q^2/U_{\rm min}^4}+
{q^2 \over U_{\rm min}^4}
\ln\left(r+\sqrt{r^2-q^2/U_{\rm min}^4}\right)\right]_{L_1}^{L_2}.
\label{analytic3}
\eea
Using (\ref{pp}) in this we obtain 
\beq
S\approx \frac{1}{2}U_{\rm min}^2R^2\left( (L_2^2-L_1^2)
+\frac{h^2}{\ln (L_2/L_1)}
\right)\approx\frac{1}{2}b^2\left( (R_2^2-R_1^2)
+\frac{H^2}{\ln (L_2/L_1)}\right).
\eeq
Therefore exp$(-S)$ is a Gaussian in $h$
with a logarithmic width, as we have seen from the 
numerical results.  The above expression for $S$ is identical
to the flat space result, eqs.\ \rf{Sflat} and \rf{d2}, with
$\s$ replaced by the AdS/CFT string tension
in eq.\ \rf{s_ads}.

\FIGURE[h]{
\begin{tabular}{cc}
\scalebox{0.28}{\rotatebox{270}{\includegraphics{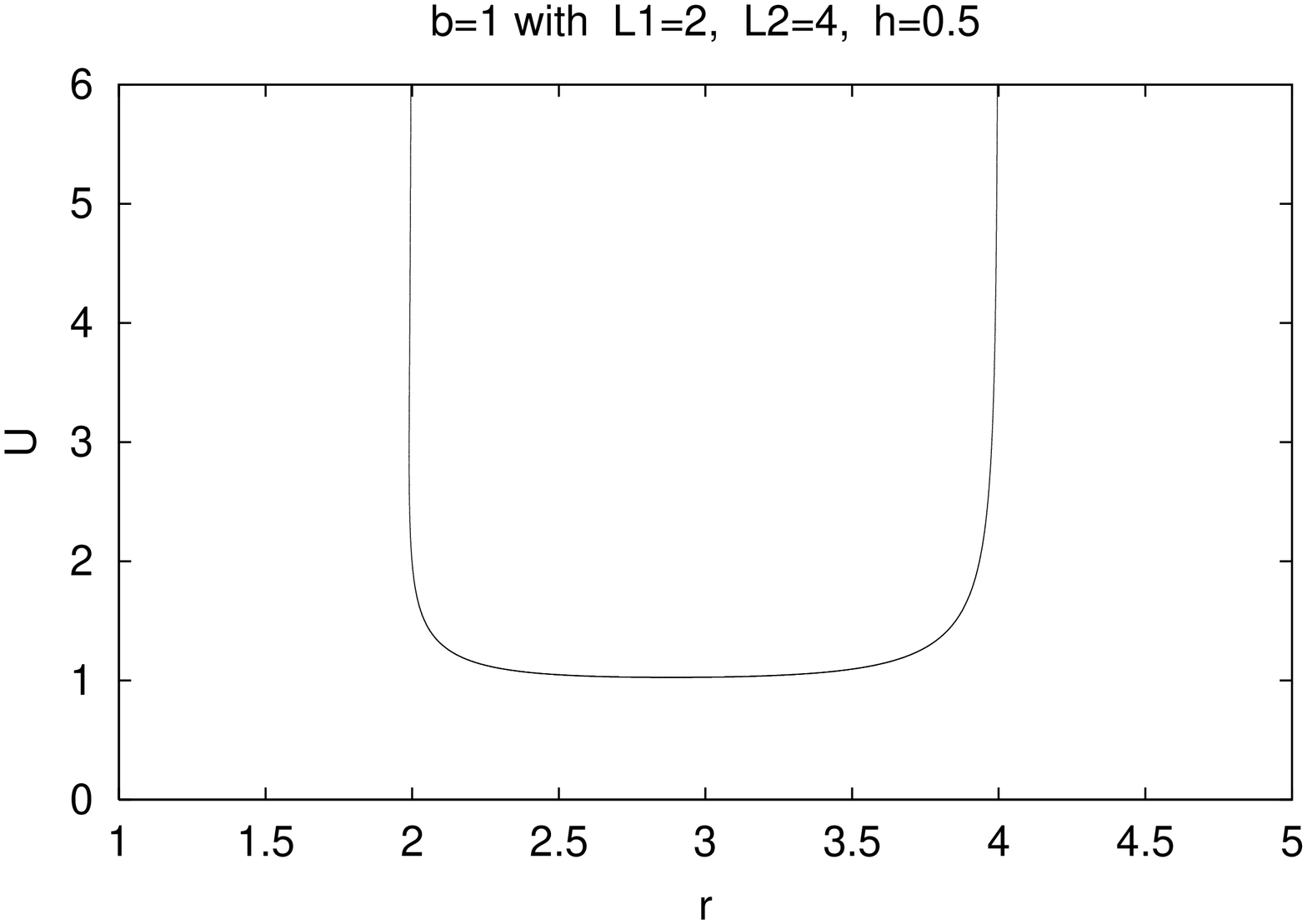}}} &
\scalebox{0.28}{\rotatebox{270}{\includegraphics{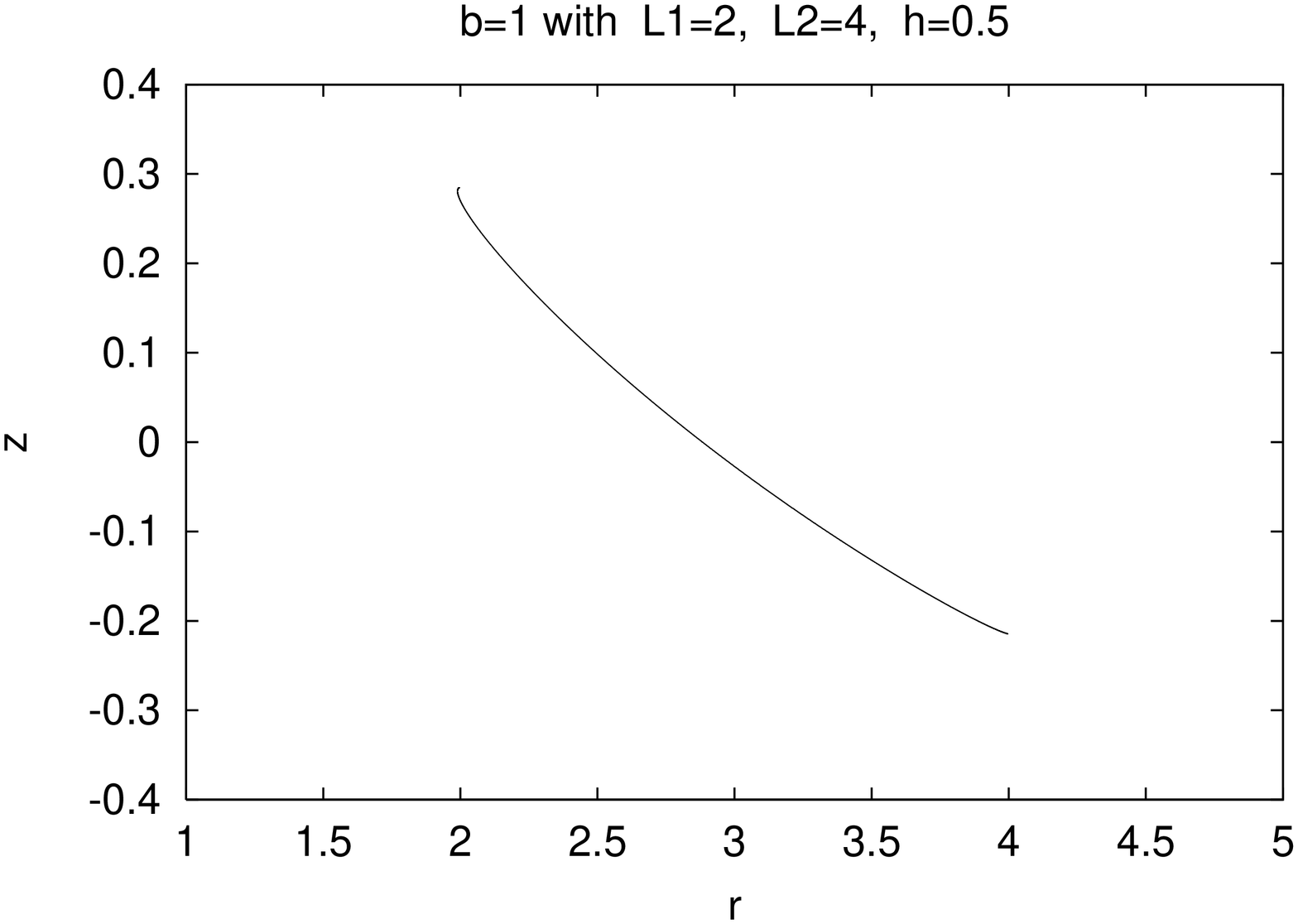}}} \\
\end{tabular}
\caption{Profiles of $U(r)$ and $z(r)$ for finite temperature.}
\label{b_finite}
}   

\FIGURE[h]{
\begin{tabular}{cc}
\scalebox{0.28}{\rotatebox{270}{\includegraphics{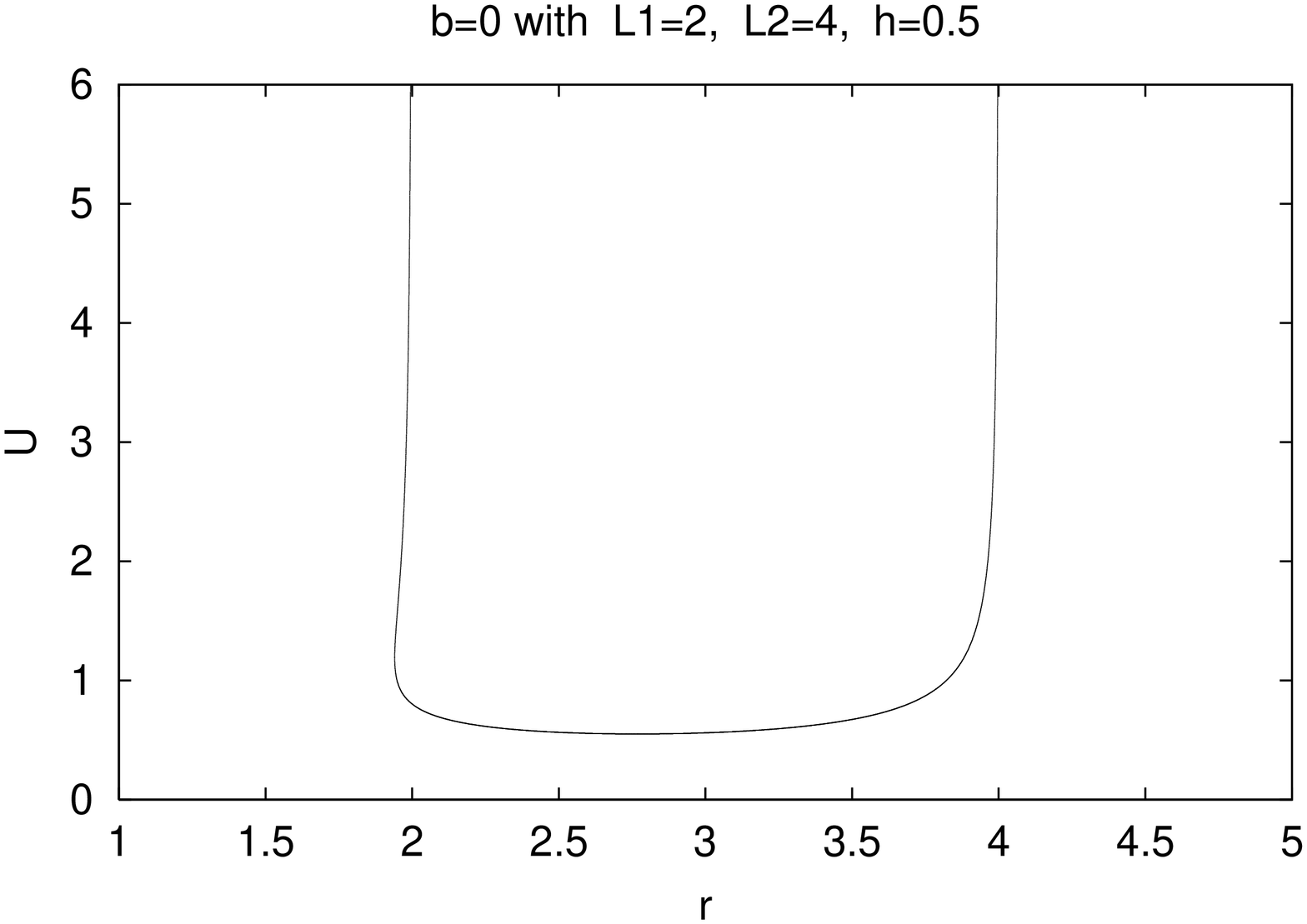}}} &
\scalebox{0.28}{\rotatebox{270}{\includegraphics{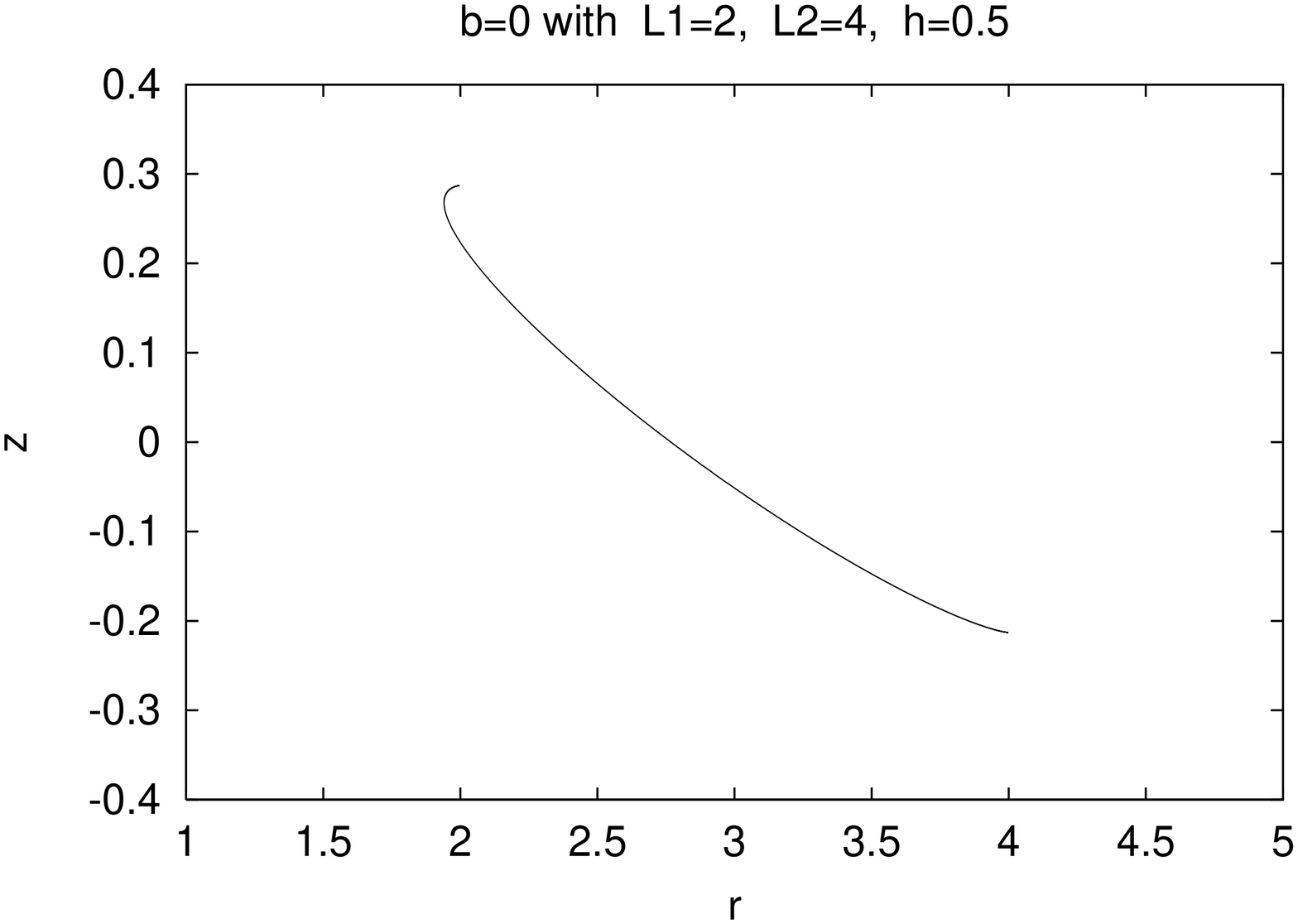}}} \\
\end{tabular}
\caption{Profiles of $U(r)$ and $z(r)$ for zero temperature. Otherwise
the parameters are the same as used in Fig.\ \ref{b_finite} in the
finite temperature case.}
\label{b_zero}
}

In Figs.\ \ref{b_finite} and \ \ref{b_zero} we have shown the behaviors
of $U$ in two cases with the same    
$L_1$, $L_2$ and $h$, but in Fig.\ \ref{b_finite} the 
temperature is finite, whereas it is zero in Fig.\ \ref{b_zero}. 
It is seen that the two cases superficially look
very similar. In spite of this, there is only logarithmic broadening
in the first case. Thus the AdS/CFT correspondence is able, by subtle effects,
to distinguish the two cases.

If we turn to the approximations given in Eqs. 
(\ref{analytic1})-(\ref{analytic3}) it 
is easy to see that they break down in the zero temperature case: There is
still a plateau, but of course the string tension disappears:
For $b=0$ we have $U_{\rm min}^2\propto 1/L^2
\rightarrow 0$, as shown in ref. \cite{1}. The logarithmic term
on the right hand side of Eq. 
(\ref{analytic3}) is therefore multiplied by an overall factor which
vanishes in the limit $L\rightarrow\infty$. Since the derivation of
the analytic approximation is only asymptotic, subdominant terms have already
been disregarded, and hence the term of order $1/L_1^2\ln (L_2/L_1)$ should
also be disregarded. We therefore have the satisfactory situation that
in the supersymmetric case the logarithmic broadening disappears.

\section{Four dimensional QCD}

So far we have discussed $QCD_3$. However it is not difficult to extend
the method to four dimensional $QCD$. Making rescalings similar to the 
previous ones we obtain the action
\beq
S\propto\int dr ~ rU^3\sqrt{1+z'^2+\frac{U'^2}{U^3-b^2}}.
\eeq
Here a prime denotes differentiation with respect to $r$.
Proceeding as before we get the equations of motion
\beq
z''=-\frac{rU^6}{q^2}z'^3,
\eeq
where $q$ is again a constant of integration, and 
\beq
U'^2=(U^3-b^3)\left(\frac{r^2U^6}{q^2}z'^2-z'^2-1\right).
\eeq
These equations are very similar to the three dimensional ones. We have
checked that the profiles for $U$ and $z$ are extremely similar in
$QCD_3$ and $QCD_4$, with plateaus in the $U(r)$ profiles. The analytic 
approximation in the previous section can therefore be performed as before, 
leading again to a logarithmic broadening of the string.
It is easy to show that the asymptotic result \rf{asym2} remains valid,
while eq.\ \rf{asym1} is replaced by
\beq
     U \approx \sqrt{{2\over 5L}}{1 \over L-r}
\eeq
leading to a somewhat faster approach to infinity at the end points.

\section{Conclusions}

   We have studied the broadening of the $QCD_3$ flux tube,
as probed by a Wilson loop whose extension is small compared
to the quark separation, in
the AdS/CFT correspondence at strong couplings.
It is found that the cross-sectional area of flux tube, as measured
by the loop probe, grows logarithmically with quark separation,
in the manner first suggested in ref.\ \cite{l}.  Thus there is no
roughening phase transition, as exists in lattice gauge theory, to 
frustrate the extrapolation of strong coupling results to weaker
couplings.  Possibly this extrapolation could be carried out by
resummation methods.  We have also found that in the zero-temperature limit,
our results are consistent with the absence of a flux tube, and no
logarithmic broadening of the field-strength distribution in a transverse
direction.

   We have not yet answered the question: ``How thick are chromo-electric
flux tubes?''  This is because the thickness, as probed by a small
Wilson loop, also depends logarithmically on the radius of the small
loop, as seen in eq.\ \rf{d}.  However, the finite temperature
formulation only resembles $QCD_3$ at length scales greater than
the inverse temperature, which serves as a kind of short distance cutoff.
Questions pertaining to $QCD_3$ can only be addressed, in the 
finite-temperature AdS/CFT correspondence, in terms of observables
whose relevant length scales are larger than this effective cutoff.
From this point of view the smallest loop probe that we can use,
and still be talking about $QCD_3$, would have a radius $R_1\equiv RL_1$ 
on the order of the inverse temperature $R_1 \sim 1/T = \pi R/b$,
in which case the diameter of the flux tube associated with a 
circular loop of radius $R_2$ would be
\beq
        d \approx {1\over b} \left[ 2 
           \ln\left({bR_2 \over \pi R}\right) \right]^{1/2}
\eeq
    
   Of course, whether or not it is relevant to $QCD_3$, the behavior
of $d$ in the $L_1 \ra 0$ limit is well defined in the AdS/CFT
correspondence.  In this limit the saddle-point approximation that we
have been using certainly breaks down, and it is necessary to compute
the full off-shell string propagator in $AdS_5\times S_5$ between a
large loop and a second loop shrunk to a point.  
It would be interesting to have further information about the
behavior of $d$ in this limit.

\acknowledgments{J.G.\ acknowledges the financial support of
MaPhySto, Centre for Mathematical Physics and Stochastics, funded by
the Danish National Research Foundation.
J.G.'s research is also supported in part by the
U.S. Department of Energy under Grant No.\ DE-FG03-92ER40711.}

\end{document}